\documentclass[12pt]{article}

\usepackage{graphicx}

\usepackage[utf8]{inputenc}
\usepackage[T1]{fontenc}
\usepackage{enumitem}
\usepackage{amsmath, amsthm, amssymb,scalerel}
\usepackage{mathtools}
\usepackage[linesnumbered,boxruled,ruled,noline]{algorithm2e}
\usepackage{algpseudocode}
\usepackage{amsfonts}
\usepackage{graphicx}
\usepackage{psfrag}
\usepackage{float}
\usepackage{caption}
\usepackage{mathbbol}
\usepackage{subcaption}
\usepackage{dsfont}

\usepackage{authblk}
\usepackage[hidelinks]{hyperref}
\usepackage[labelfont=bf]{caption}
\usepackage[symbol]{footmisc}

\usepackage[labelformat=simple]{subcaption}

\usepackage{color}
\usepackage{booktabs}
\usepackage[dvipsnames]{xcolor}
\usepackage{multirow}
\usepackage{stackrel}
\numberwithin{equation}{section}
\usepackage{caption}
\def\barroman#1{\sbox0{#1}\dimen0=\dimexpr\wd0+1pt\relax
  \makebox[\dimen0]{\rlap{\vrule width\dimen0 height 0.06ex depth 0.06ex}%
    \rlap{\vrule width\dimen0 height\dimexpr\ht0+0.03ex\relax 
            depth\dimexpr-\ht0+0.09ex\relax}%
    \kern.5pt#1\kern.5pt}}

\SetKwInput{KwInput}{Input}                
\SetKwInput{KwOutput}{Output}
\SetKwInput{tol}{tol}                
\SetKwBlock{Loop}{Loop}{end}

\SetCommentSty{mycommfont}

\usepackage{romannum}

\usepackage[a4paper, total={6in, 9in}]{geometry}

\usepackage[backend=biber,sorting=nyt,style=authoryear,natbib=true,maxcitenames=1,maxbibnames=99,doi=false,isbn=false,url=false,eprint=false,giveninits=true,date=year]{biblatex}
\AtEveryBibitem{\clearlist{language}}
\AtEveryBibitem{\clearfield{note}}
\bibliography{lit_zotero.bib}

\newcommand{\blds}[1]{\mbox{\scriptsize \boldmath $#1$}}
\newcommand{\rmd}{{\,\rm  d}}

\newcommand{\T}[1]{{#1}^{\sf  T}}

\newcommand{\z}[1]{{\tilde{#1}}}

\newcommand{\norm}[1]{| #1 |}
\newcommand{\normg}[1]{\left\Vert \, #1 \, \right\Vert}

\newcommand{\grad}[1]{{\rm grad}\left( #1 \right)}
\renewcommand{\div}[1]{{\rm div }\left( #1 \right)}

\newcommand{\la}{\langle}
\newcommand{\ra}{\rangle}

\newcommand{\fempty}[1]{{}}

\newcommand{\sty}[1]{\mbox{\boldmath $#1$}}

\newcommand{\fn}{\sty{ n}}

\newcommand{\fx}{\sty{ x}}

\newcommand{\fB}{\sty{ B}}

\newcommand{\fH}{\sty{ H}}

\newcommand{\zfH}{\tilde{\sty{ H}}}

\newcommand{\uR}{\un{R}}

\newcommand{\bfB}{\bar{\sty{ B}}}

\newcommand{\bfH}{\bar{\sty{ H}}}

\newcommand{\uxi}{\un{\xi }}

\newcommand{\cC}{{\cal C}}

\newcommand{\cfH}{\sty{{\cal H}}}

\newcommand{\zphi}{\tilde{\mbox{$\varphi $}}}

\newcommand{\un}[1]{\underline{#1}}

\title{Computational Homogenization in 3D Magnetostatics using E3C Hyper-Reduction}
\author{Hauke Goldbeck$^{1*}$, Stephan Wulfinghoff$^{1}$}
\date{$^{1}$Computational Materials Science, Department of Materials Science, Kiel University, Kaiserstr.~2, 24143 Kiel, Germany\\
hago/swu@tf.uni-kiel.de \\ [2ex]
\today}

\begin{document}
  \maketitle
  \pagenumbering{arabic}
  
  \begin{abstract}
    The recently published hyper-reduction method "Empirically Corrected Cluster Cubature" (E3C) is for the first time applied in three dimensions (here magnetostatics). The method is verified to give accurate results even for a small number of integration points, such as 15 for 3D microstructure simulations. The influence of the number of snapshots and modes, as well as the number of integration points, is investigated and the set with the best performance is selected, showing hyper-reduction errors of less than 1\%. Exemplary simulations, including a two-scale simulation are considered illustrating the performance of the E3C method for 3D simulations.

  \end{abstract}

  \section{Introduction}
\begin{figure}[h]
      \begin{center}
      \includegraphics[width=\textwidth]{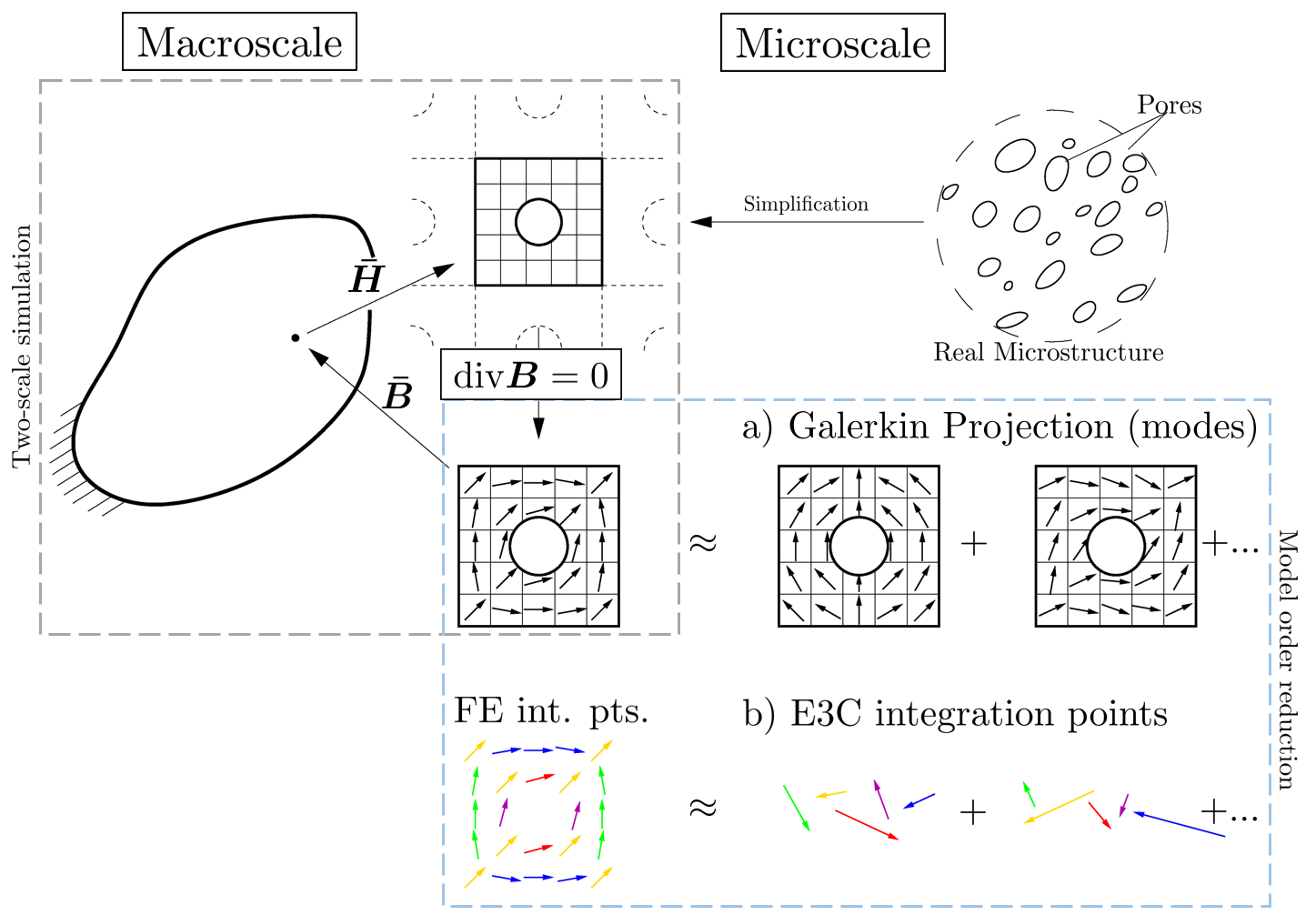}
      \end{center}
      \caption{Schematic of two-scale simulation with model order reduction.}
      \label{MicroMacroScheme}
    \end{figure}

    To efficiently solve multiscale problems it is important to improve the performance of methods like FE$^2$ (described in, e.g., \cite{kouznetsova_approach_2001,schroder_numerical_2014}) by developing faster model order reduction approaches. In general, reduced order models approximate large systems by smaller ones, thus reducing the cost of evaluating the system. The main goal is to obtain similar accuracy when solving the smaller system compared to the original large system. In recent years, several approaches to model order reduction have been developed, and in this paper we will focus on projection-based methods, specifically the Galerkin projection method (described in e.g. \cite{dar_reduced_2023}). They all have in common that a lower dimensional space can be used to represent fields originally defined in large dimensional spaces. In a preceding offline step, a lower dimensional space can be found using methods such as proper orthogonal decomposition (POD), which is generally described in \citet{chatterjee_introduction_2000} and applied in an early two-scale method in \citet{yvonnet_reduced_2007}. This offline step only needs to be done once. Afterwards, the online computation can be performed repeatedly at lower computational cost. This shifts the bottleneck from solving the global system of equations to the material law at each integration point. To further reduce the computational effort, "hyper-reduction" approaches are considered within the computational homogenization approaches (\cite{ryckelynck_hyper-reduction_2009,ares_de_parga_hyper-reduction_2023}). They further reduce the computational cost associated with evaluating the material law at each integration point. A particularly robust example is the Empirical Cubature Method (ECM) (\cite{hernandez_dimensional_2017, lange_monolithic_2024}), which allows to generate a subset of the integration points of the full order model while preserving the properties of the reduced order model as accurately as possible. In this paper, the Empirically Corrected Cluster Cubature (E3C) method is used, which was recently proposed by \citet{wulfinghoff_empirically_2025} and successfully applied in \citet{wulfinghoff_e3c_2025} and \citet{wulfinghoff_homogenization_2025}. The difference between E3C and ECM lies in the fact that E3C does not select the new set of integration points from the existing set of the finite element model, but defines a new set based on clustering and optimization techniques (compare \cite{wulfinghoff_model_2018}), which is in part motivated from statistically compatible hyper-reduction (\cite{wulfinghoff_statistically_2024}), which defines generalized integration points (IPs) in strain space. The novelty of this contribution lies in the performance investigation of the method in three dimensions, here for magnetostatic two-scale problems. The general approach is illustrated schematically in Fig.~\ref{MicroMacroScheme}. The left-hand side shows the macroscale with a body, the real microstructure of which is shown in the upper right corner. The microstructure is simplified to be used as a periodic reduced volume element at each material point of the macroscopic structure. The two-scale model connects the macro- and microscale by passing the macroscopic H-field ($\bfH$) to the microscale, where the microscopic response is calculated according to the local material and Gauss’ laws. The resulting volumetric average B-field ($\bfB$) is passed back to the macroscale as solution of the micro simulation. In order to speed up two-scale simulations, a model order reduction approach is introduced on the microscale, which is shown at the bottom right of Fig.~\ref{MicroMacroScheme}, being split into a) Galerkin projection and b) E3C method. The Galerkin projection represents the H-field by several different modes as shown in a). The hyper-reduction is introduced in b), with clustering of similar H-fields of the finite element (FE) integration points (indicated by different colored arrows). As the H-field can now be evaluated only for the cluster averages (cluster centers), the computational cost is drastically reduced compared to evaluating the material response at all FE integration points. The E3C integration points further correct the clusters, such that the average microscopic response matches the fully integrated reduced order models response as accurate as possible. In the following, the main details of the E3C method presented in \citet{wulfinghoff_empirically_2025} are repeated for the convenience of the reader. Finally, exemplary simulation results for the reduced order model, the hyper-reduced clustered model and the hyper-reduced E3C model are compared in terms of computational cost and accuracy.

  \section{Microscopic boundary value problem}
    To simulate the magnetostatic behaviour of a microstructure, Gauss' law is considered in its strong form as follows
    \begin{equation}
      \div{\fB(\fx)}=0
     \end{equation}
    with $\fB(\fx)$ being the position dependent magnetic flux density. The corresponding unknown magentic field strength $\fH(\fx)$ is introduced as
    \begin{equation}
      \fH(\fx) = \bfH + \zfH(\fx) = \bfH - \grad{\zphi(\fx)},
    \end{equation}
    with $\bfH$ representing the macroscopic magnetic field strength, $\zfH(\fx)$ is the position dependent fluctuation and $\zphi(\fx)$ denotes the fluctuation of the magnetic potential, which is set to be the primary unknown throughout this contribution. The corresponding constitutive relation of $\fB$ and $\fH$ is assumed to be given by
    \begin{equation}
      \fB = \fB(\fx,\fH),
      \label{EQConstitutiveRelation}
    \end{equation}
    where the position dependence is due to the heterogeneity of the microstructure. The resulting weak form of Gauss' law reads
    \begin{equation}
      \int_{\Omega}\delta\zfH \cdot \fB \rmd\Omega = 0
      \label{EQWeakFormFEM}
    \end{equation}
    with $\Omega$ describing the periodic microstructure and $\delta\zfH$ the variation of $\zfH$ as
    \begin{equation}
      \delta\zfH = -\grad{\delta\zphi},
    \end{equation}
    with $\delta\zphi$ assumed periodic.

  \section{Reduced order model}
    To improve the efficiency and computational cost of the microstructural boundary value problem (BVP) a reduced order model (ROM) is considered. The primary unknown $\zphi$ is now described as
    \begin{equation}
      \zphi(\fx)=\sum_{k=1}^{N_{{\rm md}}}\xi_{k}\z{\Phi}_{k}(\fx),
      \label{EQFluctuationModes}
    \end{equation}
    with $\z{\Phi}_{k}$ defining the fluctuation-modes identified by proper orthogonal decomposition~(POD) and $\xi_{k}$ the unknown mode coefficients. The number of modes considered is~$N_{{\rm md}}$. In analogy, $\fH(\fx)$ can be represented as
    \begin{equation}
      \fH(\fx) = \bfH +\underbrace{\sum_{k=1}^{N_{{\rm md}}}\xi_{k}\z\cfH_{k}(\fx)}_{\blds{\z\fH(\fx)}}\quad{\rm with}\quad\z\cfH_{k}(\fx)=-\grad{\z{\Phi}_{k}(\fx)}
      \label{EQHModesROM}
    \end{equation}
    with $\z\cfH_{k}$ describing the H-modes. Using $N_{{\rm IP}}^{{\rm FE}}$ finite element integration points with position vectors $\fx^{p}\, (p\in\{1,...,N_{{\rm IP}}^{{\rm FE}}\})$, the Galerkin projected weak form of the microscopic BVP can be derived from Eq.~\ref{EQWeakFormFEM} and Eq.~\ref{EQHModesROM}. The resulting residual vector~$\uR\,\in\mathds{R}^{N_{{\rm md}}}$ has the components
    \begin{equation}
      R_{k}=\sum_{p=1}^{N_{{\rm IP}}^{{\rm FE}}}\z\cfH_{k}(\fx^{p})\cdot \fB(\fx^{p},\fH(\fx^{p}))\Omega_{{\rm FE}}^{p}
      \label{EQResidualROM}
    \end{equation}
    with the domains of the finite element integration points $\Omega_{{\rm FE}}^{p}$, and must vanish:
    \begin{equation}
      \uR=\T{(R_{1}, ..., R_{N_{{\rm md}}})}=\underline{0}.
      \label{EQResidualROM}
    \end{equation}
    The unknown mode coefficients $\uxi$ can now be calculated by solving the above nonlinear equation system for a given $\bfH$. With the microscopic BVP solved, the material response according to the constitutive relation can be described by
    \begin{equation}
      \bfB=\la\fB\ra=\frac{1}{\Omega}\int_{\Omega}\fB\rmd\Omega\approx\frac{1}{\Omega}\sum_{p=1}^{N_{{\rm IP}}^{{\rm FE}}}\fB(\fx^{p},\fH(\fx^{p}))\Omega_{{\rm FE}}^{p},
      \label{EQBMacroROM}
    \end{equation}
    with $\la\fB\ra$ being the volume average of $\fB$ throughout the whole periodic microstructure and $\bfB$ denoting the macroscopic material response.

  \section{Hyper-reduction}
  \subsection{K-means clustering}
  \label{SectionKMeans}
    In order to achieve even more efficient calculations of the microscopic BVP, the integration point number is drastically reduced to $N_{{\rm IP}}^{{\rm HR}} \ll N_{{\rm IP}}^{{\rm FE}}$. The hyper-reduced (HR) integration points (IP) $N_{{\rm IP}}^{{\rm HR}}$ are identified by a clustering algorithm individually for each phase. In this contribution a k-means clustering approach  (\cite{macqueen_methods_1967}) is considered. Integration points exhibiting similar magnetic field strengths $\fH$ are clustered (compare, e.g., \cite{cavaliere_efficient_2020}) and the cluster centers $\cC^{q}$ are chosen to be the new integration points for the hyper-reduced approach, with each finite element integration point being part of one cluster. This is described by
    \begin{equation}
      \z\cfH^{q}=(\z\cfH_{1}^{q},...,\z\cfH_{N_{{\rm md}}}^{q}) \in \mathds{R}^{d\cdot N_{{\rm md}}}\quad(q=1,...,N_{{\rm IP}}^{{\rm HR}})
    \end{equation}
    and
    \begin{equation}
      \z\cfH^{q}=\frac{1}{\Omega^{q}}\sum_{p\in\cC^{q}}\z\cfH(\fx^{p})\Omega_{{\rm FE}}^{p} \quad {\rm with} \quad \Omega^{q}=\sum_{p\in\cC^{q}}\Omega_{{\rm FE}}^{p},
    \end{equation}
    where each finite element integration point is considered with its individual weight~$\Omega_{{\rm FE}}^{p}$. The finite element cluster average of the microscopic field strength $\fH$ is thus exactly conserved:
    \begin{equation}
      \fH^{q}:=\bfH+\sum_{k=1}^{N_{{\rm md}}}\xi_{k}\z\cfH_{k}^{q}=\frac{1}{\Omega^{q}}\sum_{p\in\cC^{q}}\fH(\fx^{p})\Omega_{{\rm FE}}^{p}.
    \end{equation}
    With the hyper-reduced integration points at hand, the equations to be solved (Eqns.~\ref{EQResidualROM} \&~\ref{EQBMacroROM}) change to
    \begin{equation}
      R_{k}\approx\sum_{q=1}^{N_{{\rm IP}}^{{\rm HR}}}\z\cfH_{k}^{q}\cdot\fB^{q}\Omega^{q}=0
      \label{EQResidualHR}
    \end{equation}
    and
    \begin{equation}
      \bfB\approx\frac{1}{\Omega}\sum_{q=1}^{N_{{\rm IP}}^{{\rm HR}}}\fB^{q}\Omega^{q}.
      \label{EQBMacroHR}
    \end{equation}
    Here $\fB^{q}$ is given by
    \begin{equation}
      \fB^{q}=\fB^{q}(\fH^{q}).
    \end{equation}

    \subsection{E3C method}
    The E3C method improves the identified hyper reduced integration points (i.e., the~$\z\cfH^{q}$) by empirically correcting them, such that the reduced order model equations (Eqns.~\ref{EQResidualHR} \&~\ref{EQBMacroHR}) are solved as accurate as possible compared to the fully integrated reduced order model. To correct the integration points, a training step is performed using the fully integrated reduced order model, which shows accurate results compared to finite elements. The results $\uxi^{s}$ and $\bfB^{s}$ for ($s\in{1,...,N_{{\rm full}}}\quad{\rm with}\quad N_{{\rm full}}=N_{{\rm sim}}\cdot N_{{\rm steps}}$, $N_{{\rm sim}}$: Number of simulations in the training data set, $N_{{\rm steps}}$: Number of time steps of each $N_{{\rm sim}}$) are collected and the following cost function  $c$
    \begin{equation}
      \begin{split}
      c(\z\cfH^{1},...,\z\cfH^{N_{{\rm IP}}^{{\rm HR}}}):=\frac{1}{2}\sum_{s=1}^{N_{{\rm full}}}\sum_{k=1}^{N_{{\rm md}}}\Biggl(\frac{1}{\Omega}\sum_{q=1}^{N_{{\rm IP}}^{{\rm HR}}}\z\cfH_{k}^{q}\cdot\fB^{q}(\fH^{qs})\Omega^{q}\Biggr)^{2} \\
      +\frac{a}{2}\sum_{s=1}^{N_{{\rm full}}}\normg{\frac{1}{\Omega}\sum_{q=1}^{N_{{\rm IP}}^{{\rm HR}}}\fB^{q}(\fH^{qs})\Omega^{q}-\bfB^{s}}^{2}
      \end{split}
      \label{ErrorEQ}
    \end{equation}
    is minimized ($a$ is a user defined weight; here $a=10^{-5})$. Here, the magnetic field strength $\fH^{qs}$ is given by
    \begin{equation}
      \fH^{qs}(\z\cfH^{q})=\bfH^{s}+\sum_{l=1}^{N_{{\rm md}}}\xi_{l}^{s}\z\cfH_{l}^{q}.
    \end{equation}
    Additionally the constraint
    \begin{equation}
      \sum_{q=1}^{N_{{\rm IP}}^{{\rm HR}}}\z\cfH^{q}\Omega^{q}=\boldsymbol{0}
    \end{equation}
    is applied by elimination of the last component $\z\cfH^{N_{{\rm IP}}^{{\rm HR}}}$. This ensures that the overall fluctuation is zero on average
    \begin{equation}
      \la\zfH\ra=\frac{1}{\Omega}\sum_{q=1}^{N_{{\rm IP}}^{{\rm HR}}}\sum_{k=1}^{N_{{\rm md}}}\xi_{k}\z\cfH_{k}^{q}\Omega^{q}=\boldsymbol{0}\quad\Leftrightarrow\quad\la\fH\ra=\bfH,
    \end{equation}
    which results in the macroscopic field strength $\bfH$ being equal to the average of the microscopic field strength $\fH$. The new empirically corrected set of hyper-reduced integration points is calculated using the Polak-Ribi\`ere version of the nonlinear conjugate gradient method (\cite{polak_note_1969}). The initial solution guess is given by the k-means integration points .

  \section{Results}
\begin{figure}
      \begin{center}
      \includegraphics[width=.9\textwidth]{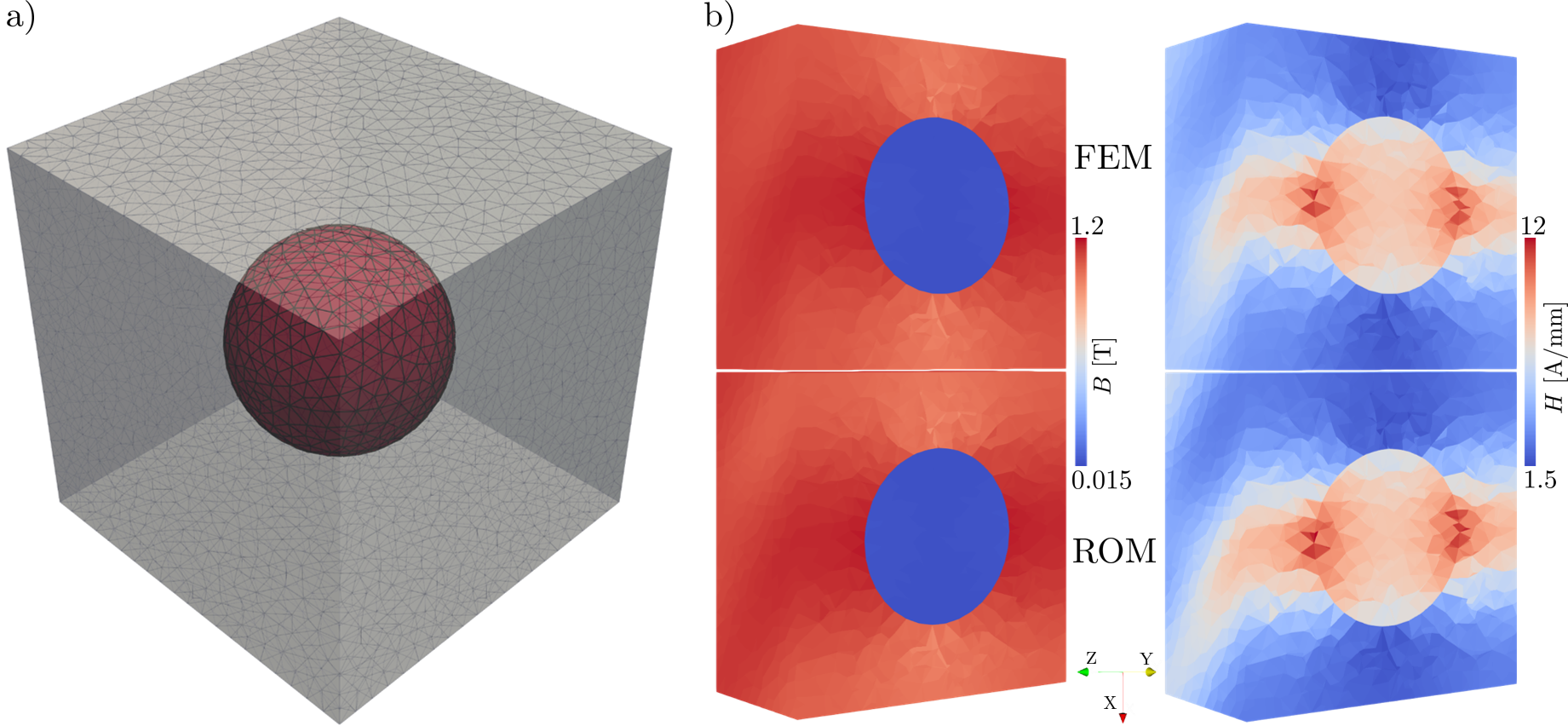}
      \end{center}
      \caption{a) Mesh used in micro simulations. b) The simulation results for a random test direction are compared with respect to the $\fB$-field and $\fH$-field of FEM and fully integrated ROM.}
      \label{MicroMesh}
    \end{figure}

    The results presented in the following chapter have been calculated using the B-H law
    \begin{equation}
      \fB=\mu_{0}\biggl[\normg{\fH}+M_{{\rm sp}}L\biggl(\frac{3\chi_{0}}{M_{{\rm sp}}}\normg{\fH}\biggr)\biggr]\frac{\fH}{\normg{\fH}}+\mu_{{\rm stab}}\fH,
      \label{EQMateriallaw}
    \end{equation}
    with $\mu_{0}$ representing the vacuum permeability, $M_{{\rm sp}}$ being the spontaneous magnetization, $\chi_{0}$ describing the initial magnetic susceptibility and $L(x)$ denoting the Langevin-function (\cite{langevin_sur_1905}). The last term involving $\mu_{{\rm stab}}$ is included to stabilize the simulation in regions of high field strength. In such regions the Langevin-function shows nearly horizontal course, which is numerically difficult to solve. The material parameters, in Tab.~\ref{TABMatPara} have been used throughout all calculations.
    \begin{table}
      \caption{Material parameters.}
      \label{TABMatPara}
    \begin{center}
    \begin{tabular}{c|c|c|c}
      material & $\chi_{0}[-]$&$\mu_{0}M_{{\rm sp}}[{\rm T}]$&$\mu_{{\rm stab}}$ \\
      \hline
      matrix &$1001$ &$1.2$ & $\mu_{0}$ \\
      \hline
      pores &$1$ &$1.2$  &$0$ \\
      \end{tabular}
    \end{center}
  \end{table}
    The phase contrast of $\sim$1000 chosen for evaluation is numerically more challenging compared to smaller phase contrasts or inverted order of materials (results are not shown). The finite element model for microscopic calculations uses a geometry (cube with edge length~1) with one spherical pore (radius=0.27; pore volume fraction of 8.24 \%) placed in the center of a cube, with the mesh shown in Fig.~\ref{MicroMesh}~a). The mesh is periodic and built from 38195 linear tetrahedrons with 1 integration point per element.
    The final macroscopic $\bfH$ is applied in the following form
    \begin{equation}
      \bfH=5\frac{M_{{\rm sp}}}{\chi_{0}}\fn
      \label{EQMacroH}
    \end{equation} 
    with $\fn$ being a normalized vector used to change the direction of applied $\bfH$ for all microscopic simulations. The simulations are done on an AMD\textsuperscript{\textregistered} Ryzen threadripper 3970X 64-core processor with 128 GB RAM. The finite element software used is FEAP 8.6 (\cite{taylor_feap_2020}) combined with the PARDISO solver (\cite{schenk_pardiso_2001}).

    \begin{figure}
      \begin{center}
        \includegraphics[width=\textwidth]{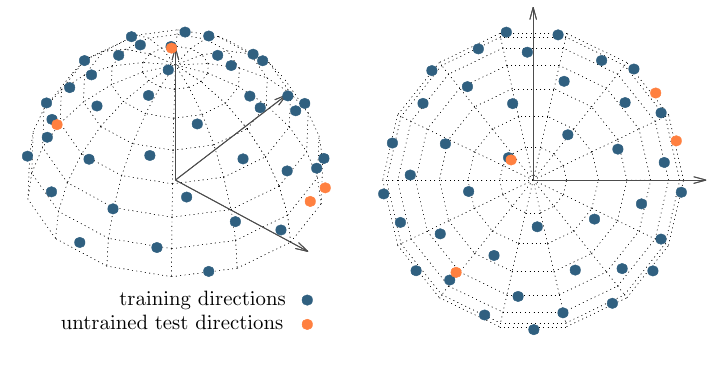}
      \end{center}
      \caption{\textbf{Left:} Half sphere with trained directions and random generated untrained test directions. \textbf{Right:} Top view of half sphere showing the distribution of trained directions from canonical Fibonacci algorithm.}
      \label{HalfSphereWithTraining}
    \end{figure}

    \subsection{Evaluation of reduced order model}
    \label{SectionEVALROM}
\begin{figure}
      \begin{center}
      \includegraphics[width=.49\textwidth]{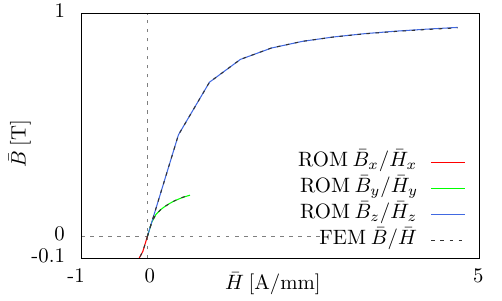}
      \includegraphics[width=.49\textwidth]{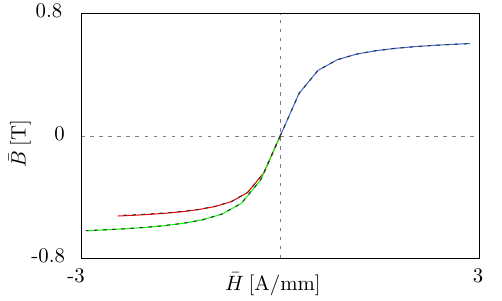}
      \includegraphics[width=.49\textwidth]{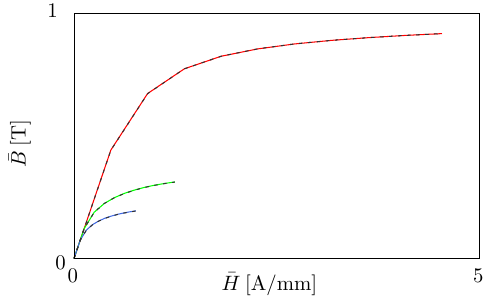}
      \includegraphics[width=.49\textwidth]{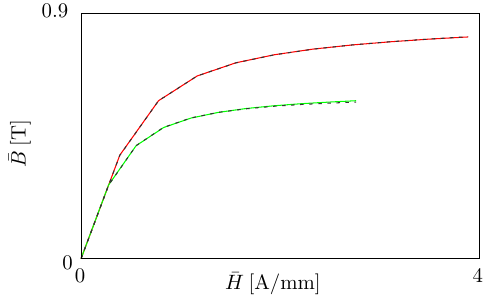}
      \end{center}
      \caption{Comparison of ROM (trained with 400 snapshots) and FEM results for the four random untrained test directions shown in Fig.~\ref{HalfSphereWithTraining}.}
      \label{ROMPerformance}
    \end{figure}

    In order to built an accurate hyper-reduced model, a fully integrated reduced order model (ROM) has to be built first. The ROM is trained from the FEM model by collecting a set of snapshots containing $N_{{\rm sim}}$ simulations of $N_{{\rm steps}}$ time steps each, which are chosen constant here. All simulations used for training differ by the direction of applied $\bfH$ (compare Eq.~\ref{EQMacroH}). This is achieved by equally distributing the $\fn$ directions on a half sphere with a canonical Fibonacci lattice algorithm (similar to \cite{roberts_evenly_2018}). The other half of the sphere is omitted due to symmetry reasons. The resulting directions $\fn$ are shown in Fig.~\ref{HalfSphereWithTraining} for the case $N_{{\rm sim}}=40$ as training directions for 400 snapshots ($N_{{\rm steps}}=10$). To check the accuracy of the ROM, four randomly generated untrained directions (shown in Fig.~\ref{HalfSphereWithTraining}) are tested. The ROM results are compared to FEM in Fig.~\ref{ROMPerformance}. The number of modes used for the fully integrated reduced order model has been chosen such that an error below 1\% (if compared with FEM) is obtained. It turns out that $N_{{\rm md}}=10$ modes are sufficient for this purpose. Increasing the number of modes would lead to a further reduction of the error, while increasing the cost for the evaluation of the ROM model. As can be observed from Fig.~\ref{ROMPerformance}, the ROM agrees very well to the FEM results for the chosen number of modes and training data. Collecting the snapshots takes $\sim$9 minutes (wall clock time). Building the ROM from the snapshots takes less than $\sim$1 minute for the POD. With a FEM simulation taking $\sim$5 s (CPU-time per time step) the speedup is already significant, as a simulation with the fully integrated ROM only takes $\sim$70 ms (CPU-time per time step).

    \subsection{Evaluation of hyper-reduction based on k-means clustering}
\begin{figure}
      \begin{center}
      \includegraphics[width=.49\textwidth]{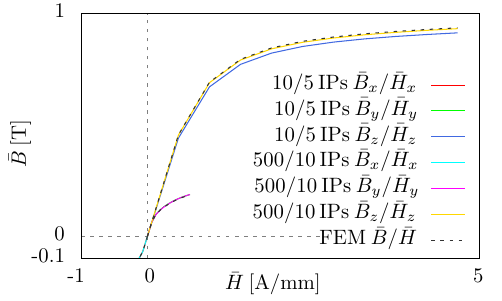}
      \includegraphics[width=.49\textwidth]{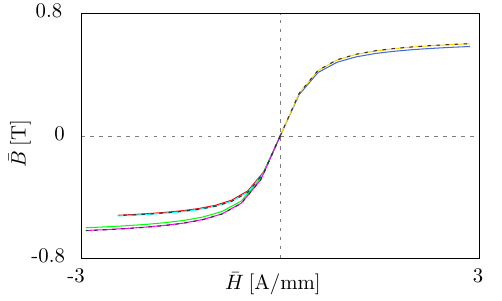}
      \includegraphics[width=.49\textwidth]{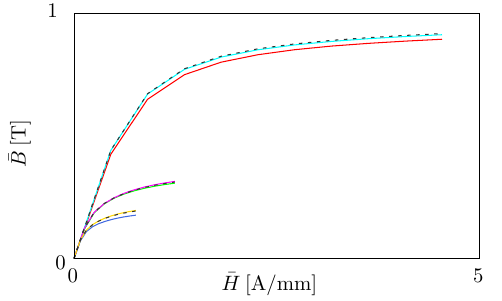}
      \includegraphics[width=.49\textwidth]{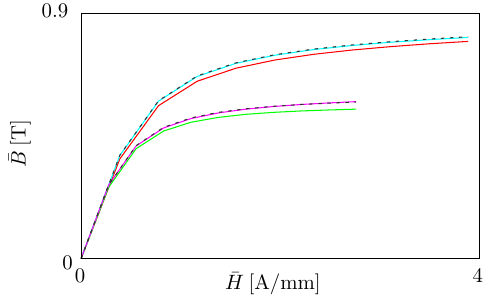}
      \end{center}
      \caption{Comparison of k-means clustering for 10/5 and 500/10 IPs with FEM results for the four random untrained test directions shown in Fig.~\ref{HalfSphereWithTraining}.}
      \label{KmeansPerformance}
    \end{figure}

    Using k-means clustering, as described in Section \ref{SectionKMeans}, the hyper-reduced (preliminary to E3C) model can be built using different numbers of integration points (IPs) for the individual phases. Aiming for a cost efficient model a low number of IPs is desirable. In Fig.~\ref{KmeansPerformance} results for two different hyper-reduced IP sets are shown in comparison with the FEM results. The two cases compared are based on 10/5 and 500/10 IPs, with the first number describing the number of integration points of the matrix material and the second number the amount of IPs for the pores. It can be observed, that the accuracy increases with a higher number of IPs. Further it is shown, that the accuracy depends less on the number of IPs of the nonmagnetic pores than on the number of IPs of the magnetic matrix material. Even for 15 overall IPs a quite good accuracy is achievable, although some error remains. The runtime changes if compared to the fully integrated ROM simulation from $\sim$70 ms to $\sim$90 $\mu$s for 15 IPs and $\sim$1 ms for 510 IPs (CPU-times per time step). 

    \subsection{Evaluation of E3C hyper-reduction}
\begin{figure}
      \includegraphics[width=\textwidth]{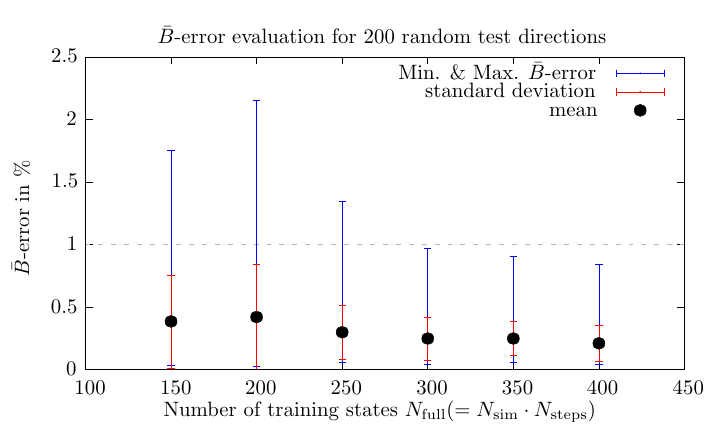}
      \caption{Comparison of trainings data amount needed for E3C method with 15 integration points to reach $\bfB$-errors below 1\%. For example, $N_{{\rm full}}=400$ was obtained through $N_{{\rm sim}}=40$ simulations with $\bfH$-directions illustrated in Fig.~\ref{HalfSphereWithTraining} with $N_{{\rm steps}}=10$ time steps each.}
      \label{E3CTrainingEvaluation}
    \end{figure}

\begin{figure}
      \begin{center}
      \includegraphics[width=.49\textwidth]{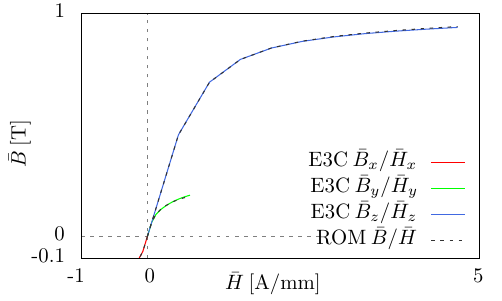}
      \includegraphics[width=.49\textwidth]{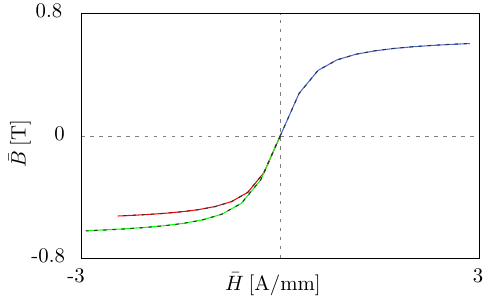}
      \includegraphics[width=.49\textwidth]{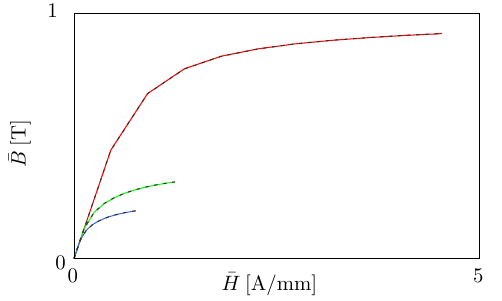}
      \includegraphics[width=.49\textwidth]{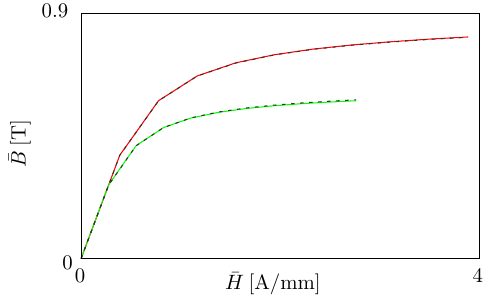}
      \end{center}
      \caption{Comparison of hyper-reduced E3C with fully integrated ROM results for 10/5 integration points and $N_{{\rm full}}=400$ for the four random untrained test directions shown in Fig.~\ref{HalfSphereWithTraining}.}
      \label{E3CPerformance}
    \end{figure}

    To minimize the cost function in Eq.~(\ref{ErrorEQ}) a set of training data has to be collected. This is done by gathering the results $\uxi^{s}$ and $\bfB^{s}$ of $N_{{\rm full}}$ fully integrated ROM simulations. $N_{{\rm full}}$ is given by $N_{{\rm sim}}$ simulations with $N_{{\rm steps}}$ constant time steps each. Trained directions $\fn$ of $\bfH$ (Eq.~\ref{EQMacroH}) are equally distributed on a hemisphere by using the same approach, as described in Section \ref{SectionEVALROM}, applying the canonical Fibonacci lattice (compare Fig.~\ref{HalfSphereWithTraining}). After training 15 E3C integration points (matrix: 10; pores: 5) with a certain dataset, the hyper-reduced ROM is tested using 200 randomly generated test directions. For each test direction a fully integrated ROM and E3C simulation are thus calculated and the relative error~E
    \begin{equation}
      E=\frac{\max\limits_{i}\norm{\hat{\bar{B}}_{i}^{{\rm full}}-\hat{\bar{B}}_{i}^{{\rm HR}}}}{\max\limits_{j}\hat{\bar{B}}_{j}^{{\rm full}}-\min\limits_{k}\hat{\bar{B}}_{k}^{{\rm full}}}\times100\%,
      \label{ErrorE3CROM}
    \end{equation} 
    is calculated, with $\hat\fB^{{\rm full}}$ and $\hat\fB^{{\rm HR}}$ being defined by
    \begin{equation}
      \begin{split}
      \hat\bfB^{{\rm full}}=(\overbrace{\la B^{{\rm full}}_{x}(t_{1})\ra,\la B^{{\rm full}}_{y}(t_{1})\ra,\la B^{{\rm full}}_{z}(t_{1})\ra}^{\blds{\bfB^{{\rm full}}(t_{1})}}, \la B^{{\rm full}}_{x}(t_{2})\ra, ... ,\la B^{{\rm full}}_{z}(N_{{\rm steps}})\ra)^\mathsf{T} \in \mathds{R}^{d \cdot N_{{\rm steps}}}\\
      \hat\bfB^{{\rm HR}}=(\underbrace{\la B^{{\rm HR}}_{x}(t_{1})\ra,\la B^{{\rm HR}}_{y}(t_{1})\ra,\la B^{{\rm HR}}_{z}(t_{1})\ra}_{\blds{\bfB^{{\rm HR}}(t_{1})}}, \la B^{{\rm HR}}_{x}(t_{2})\ra, ... ,\la B^{{\rm HR}}_{z}(N_{{\rm steps}})\ra)^\mathsf{T} \in \mathds{R}^{d \cdot N_{{\rm steps}}}\\
      \end{split}
    \end{equation}
    In other words, $\hat\bfB^{{\rm full}}$ and $\hat\bfB^{{\rm HR}}$ collect the macro-responses of both models for all time steps of a given simulation. Fig.~\ref{E3CTrainingEvaluation} shows the evaluation of relative error E and number of training data needed to achieve a desired hyper-reduction error of~<1\%. The mean value of the 200 errors calculated is depicted in Fig.~\ref{E3CTrainingEvaluation} alongside the maximum and minimum errors and the standard deviations. It can be observed that increasing the amount of training data used leads to a reduction of the overall mean error and the standard deviation as well as the maximum error. The aim is to select a training data set with errors of well below 1\%, which is fulfilled by the sets of $N_{{\rm full}}= 300,\,350\,\&\,400$. In order to use the most promising training data set, the '400' option is chosen for the following E3C calculations. In other words, the same training directions as chosen for the mode identification via POD are used for E3C training. In Fig.~\ref{E3CPerformance}, exemplary E3C results are compared with the results of the fully integrated ROM. It can be observed, that the results  match accurately. The time used to generate the training dataset with 40 simulations of 10 constant time steps takes $\sim$4 minutes (wall clock time) and the nonlinear conjugate gradient method takes $\sim$7 minutes (wall clock time) to minimize the cost function (Eq.~\ref{ErrorEQ}) without parallelization. There is no speedup compared to the clustered HR approach, but the accuracy is improved significantly by using E3C method even for a small set of IPs.

    \subsection{Two-scale simulation}
\begin{figure}[H]
      \begin{center}
        \includegraphics[width=0.8\textwidth,clip]{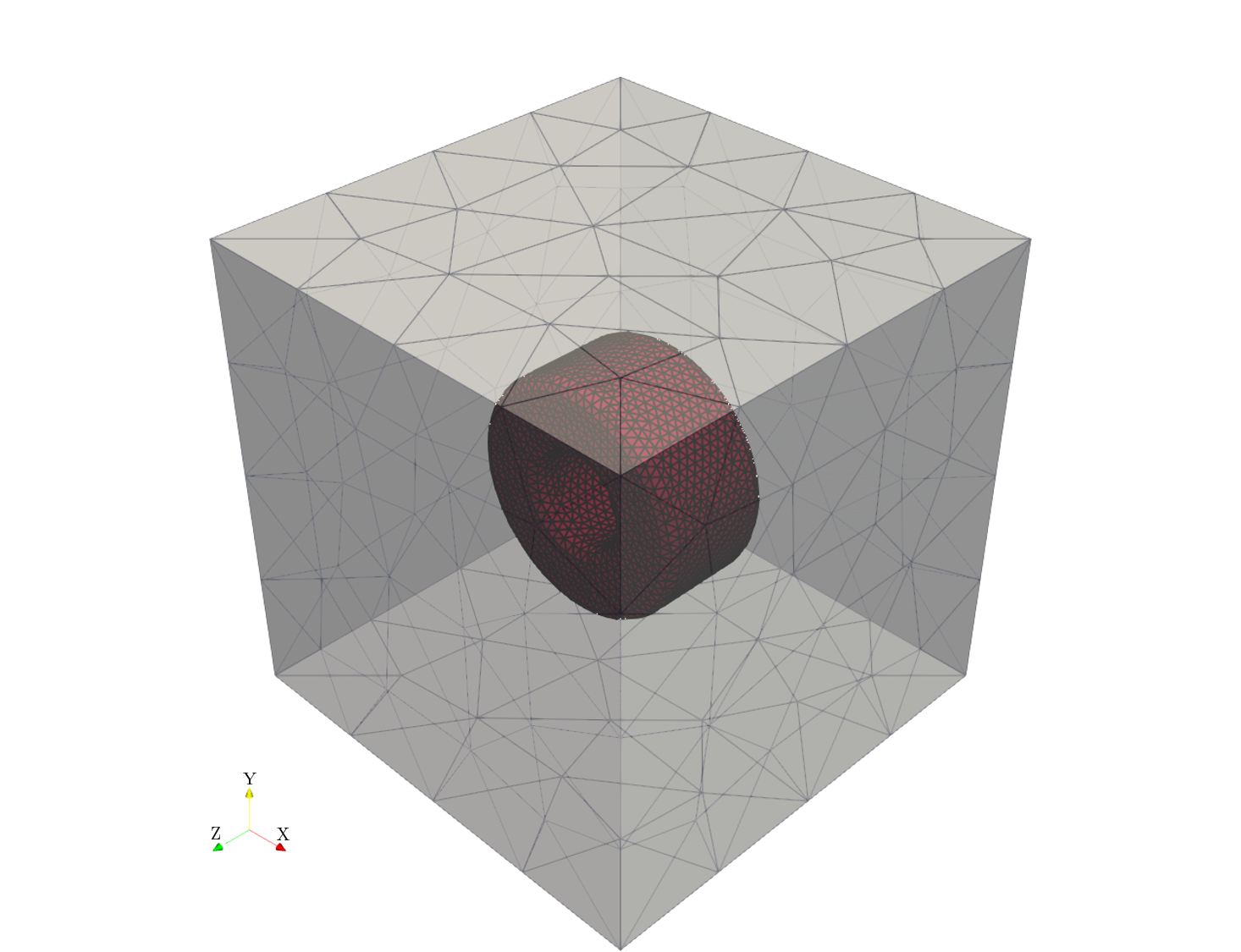}
      \end{center}
      \caption{Mesh of two-scale simulation.}
      \label{E3CMeshTwoScale}
    \end{figure}

\begin{figure}[H]
      \begin{center}
      \includegraphics[width=0.68\textwidth,clip]{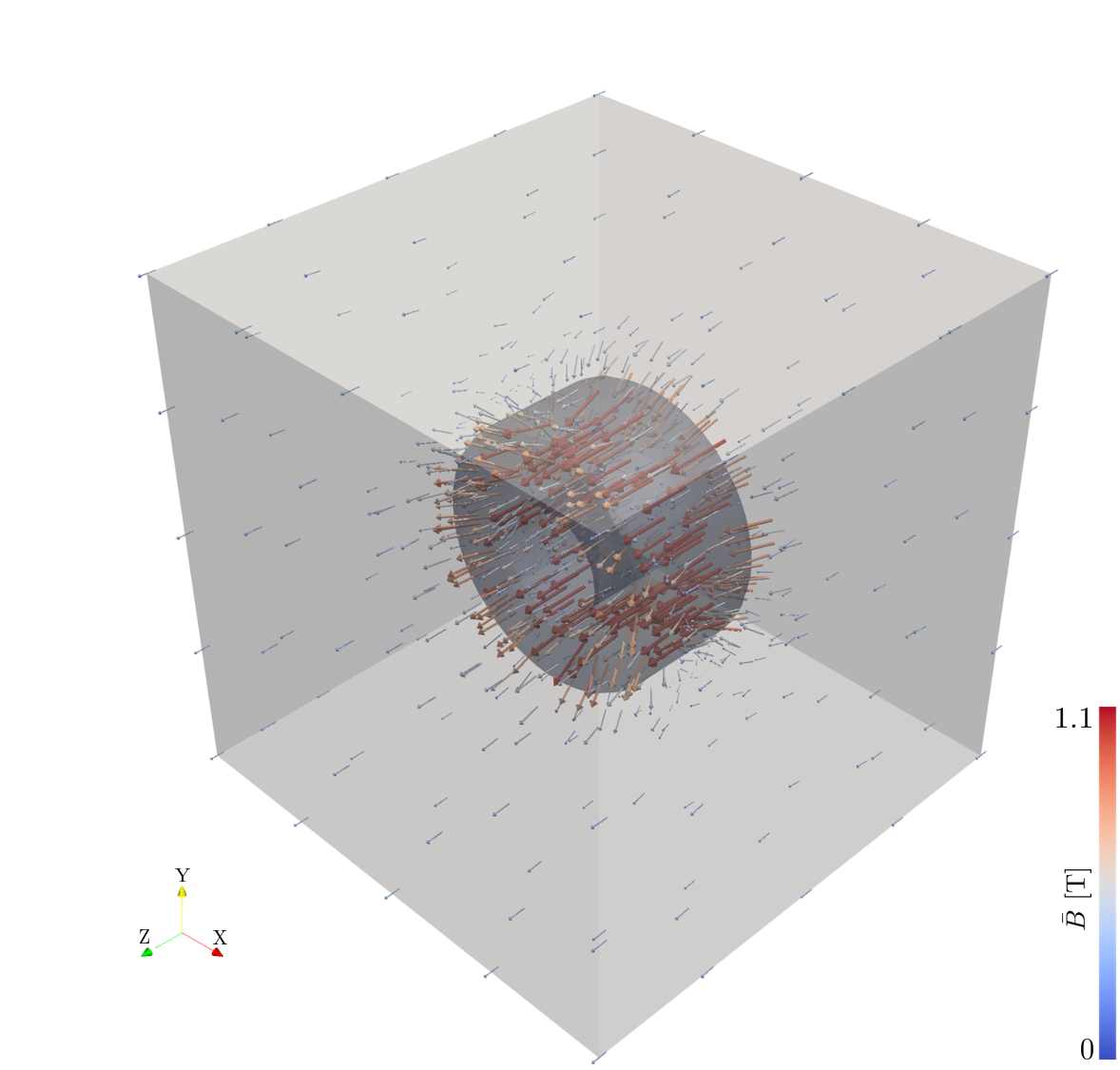}
      \end{center}
      \caption{$\bfB$-field of two-scale simulation.}
      \label{E3CTwoScaleFinal}
    \end{figure}

    The E3C method is applied in a two-scale simulation on the microstructure of a nut-like structure (covering a volume fraction of 4\%) embedded inside a free space box with an edge length of 20. The E3C-ROM is solved at each macroscopic integration point of the magnetic nut. The mesh is shown in Fig.~\ref{E3CMeshTwoScale}, with coarse linear tetrahedron elements (16525 elements) for the surrounding box and a fine mesh of linear tetrahedrons (23713 elements) for the nut structure. The macroscopic magnetic potential (+z: -5000 A; -z: 0 A) is prescribed on the two opposite surfaces (with normal in +z and -z direction) of the surrounding box and is linearly increased throughout the simulation to introduce a magnetic field. The micro geometry used here is the one shown in Fig.~\ref{MicroMesh} a). The simulation was carried out in $\sim$28~s (CPU-time per time step), using 10 time steps of equal size, while the same problem can be solved in $\sim$12~s (CPU-time per time step) for a single scale simulation. The CPU-time of $\sim$28~s per time step for the two-scale simulation includes the time for global equation system solution of $\sim$12~s (compare single scale CPU-time) as well as the solution time of the micro model at each integration point of the structure ($\sim$16~s) for multiple Newton solution steps. A parallel solver was used for the macroscopic global equation system, but not for the local equation system. The convergence of the macroscopic residual norm is depicted in Tab.~\ref{TABConvergence} for an arbitrary time step of the two-scale simulation.
    \begin{table}
      \caption{Residual norm convergence.}
      \label{TABConvergence}
    \begin{center}
      \begin{tabular}{c}
        residual norm\\
        \hline
        2.57884557E+01 \\
        1.36090642E-04 \\
        2.02831843E-06 \\
        4.60515544E-09 \\
        1.19909403E-13 \\
        8.92167174E-23 \\
        \hline
        Final: 7.91E-14 \\
        \end{tabular}
      \end{center}
    \end{table}
    The final state of the simulation can be seen in Fig.~\ref{E3CTwoScaleFinal} showing the course of the macroscopic magnetic flux density $\bfB$ using arrows.

  \section{Summary}
  The E3C method has been successfully applied to a magnetostatic two-scale model in 3D. It has been shown, that it can accurately (compared to FEM) calculate the material behavior for a phase contrast of 1000 using only 10 modes and 15 integration points for the micro model. The bottleneck of E3C method currently seems to be the nonlinear conjugate gradient method, as the time needed to built the training data and minimize the cost function is still large compared to individual solution time of one E3C calculation. The minimization should be optimized in future development of the method.

  \section*{Acknowledgements}
  This research is associated to project 286471992 (Collaborative Research Center 1261), which is supported by the German Research Foundation (Deutsche Forschungsgemeinschaft, DFG). This support is gratefully acknowledged.

  \printbibliography

@article{wulfinghoff_empirically_2025,
	title = {Empirically corrected cluster cubature ({E3C})},
	volume = {437},
	issn = {0045-7825},
	url = {https://www.sciencedirect.com/science/article/pii/S0045782525000519},
	doi = {10.1016/j.cma.2025.117779},
	urldate = {2025-03-06},
	journal = {Computer Methods in Applied Mechanics and Engineering},
	author = {Wulfinghoff, Stephan},
	month = mar,
	year = {2025},
	pages = {117779},
}

@article{macqueen_methods_1967,
	title = {Some methods for classification and analysis of multivariate observations},
	volume = {5},
	language = {en},
	journal = {Proceedings of the Fifth Berkeley Symposium on Mathematical Statistics and Probability},
	author = {MacQueen, J},
	month = jan,
	year = {1967},
	pages = {281--298},
}

@article{polak_note_1969,
	title = {Note sur la convergence de méthodes de directions conjuguées},
	volume = {3},
	issn = {0373-8000},
	url = {http://www.esaim-m2an.org/10.1051/m2an/196903R100351},
	doi = {10.1051/m2an/196903R100351},
	language = {fr},
	number = {16},
	urldate = {2025-03-06},
	journal = {Revue française d'informatique et de recherche opérationnelle. Série rouge},
	author = {Polak, E. and Ribière, G.},
	year = {1969},
	pages = {35--43},
}

@incollection{schroder_numerical_2014,
	address = {Vienna},
	title = {A numerical two-scale homogenization scheme: the {FE2}-method},
	isbn = {978-3-7091-1625-8},
	shorttitle = {A numerical two-scale homogenization scheme},
	url = {https://doi.org/10.1007/978-3-7091-1625-8_1},
	language = {en},
	urldate = {2025-03-06},
	booktitle = {Plasticity and {Beyond}: {Microstructures}, {Crystal}-{Plasticity} and {Phase} {Transitions}},
	publisher = {Springer},
	author = {Schröder, Jörg},
	editor = {Schröder, Jörg and Hackl, Klaus},
	year = {2014},
	doi = {10.1007/978-3-7091-1625-8_1},
	pages = {1--64},
}

@article{kouznetsova_approach_2001,
	title = {An approach to micro-macro modeling of heterogeneous materials},
	volume = {27},
	issn = {1432-0924},
	url = {https://doi.org/10.1007/s004660000212},
	doi = {10.1007/s004660000212},
	language = {en},
	number = {1},
	urldate = {2025-03-06},
	journal = {Computational Mechanics},
	author = {Kouznetsova, V. and Brekelmans, W. A. M. and Baaijens, F. P. T.},
	month = jan,
	year = {2001},
	pages = {37--48},
}

@article{yvonnet_reduced_2007,
	title = {The reduced model multiscale method ({R3M}) for the non-linear homogenization of hyperelastic media at finite strains},
	volume = {223},
	issn = {0021-9991},
	url = {https://www.sciencedirect.com/science/article/pii/S0021999106004402},
	doi = {10.1016/j.jcp.2006.09.019},
	number = {1},
	urldate = {2025-03-06},
	journal = {Journal of Computational Physics},
	author = {Yvonnet, J. and He, Q. -C.},
	month = apr,
	year = {2007},
	pages = {341--368},
}

@article{ryckelynck_hyper-reduction_2009,
	title = {Hyper-reduction of mechanical models involving internal variables},
	volume = {77},
	copyright = {Copyright © 2008 John Wiley \& Sons, Ltd.},
	issn = {1097-0207},
	url = {https://onlinelibrary.wiley.com/doi/abs/10.1002/nme.2406},
	doi = {10.1002/nme.2406},
	language = {en},
	number = {1},
	urldate = {2025-03-06},
	journal = {International Journal for Numerical Methods in Engineering},
	author = {Ryckelynck, D.},
	year = {2009},
	note = {\_eprint: https://onlinelibrary.wiley.com/doi/pdf/10.1002/nme.2406},
	pages = {75--89},
}

@article{hernandez_dimensional_2017,
	title = {Dimensional hyper-reduction of nonlinear finite element models via empirical cubature},
	volume = {313},
	issn = {0045-7825},
	url = {https://www.sciencedirect.com/science/article/pii/S004578251631355X},
	doi = {10.1016/j.cma.2016.10.022},
	urldate = {2025-03-06},
	journal = {Computer Methods in Applied Mechanics and Engineering},
	author = {Hernández, J. A. and Caicedo, M. A. and Ferrer, A.},
	month = jan,
	year = {2017},
	pages = {687--722},
}

@article{ares_de_parga_hyper-reduction_2023,
	title = {Hyper-reduction for {Petrov}–{Galerkin} reduced order models},
	volume = {416},
	issn = {0045-7825},
	url = {https://www.sciencedirect.com/science/article/pii/S004578252300422X},
	doi = {10.1016/j.cma.2023.116298},
	urldate = {2025-03-06},
	journal = {Computer Methods in Applied Mechanics and Engineering},
	author = {Ares de Parga, S. and Bravo, J. R. and Hernández, J. A. and Zorrilla, R. and Rossi, R.},
	month = nov,
	year = {2023},
	pages = {116298},
}

@article{lange_monolithic_2024,
	title = {A monolithic hyper {ROM} {FE2} method with clustered training at finite deformations},
	volume = {418},
	issn = {0045-7825},
	url = {https://www.sciencedirect.com/science/article/pii/S0045782523006461},
	doi = {10.1016/j.cma.2023.116522},
	urldate = {2025-03-06},
	journal = {Computer Methods in Applied Mechanics and Engineering},
	author = {Lange, Nils and Hütter, Geralf and Kiefer, Bjoern},
	month = jan,
	year = {2024},
	pages = {116522},
}

@article{taylor_feap_2020,
	title = {{FEAP} - - {A} {Finite} {Element} {Analysis} {Program}},
	language = {en},
	author = {Taylor, Robert L},
	year = {2020},
}

@misc{roberts_evenly_2018,
	title = {Evenly distributing points on a sphere},
	url = {https://extremelearning.com.au/evenly-distributing-points-on-a-sphere/},
	language = {en-US},
	urldate = {2025-03-07},
	journal = {Extreme Learning},
	author = {Roberts, Martin},
	month = aug,
	year = {2018},
}

@incollection{dar_reduced_2023,
	address = {Cham},
	title = {Reduced {Order} {Modeling}},
	isbn = {978-3-031-36644-4},
	url = {https://doi.org/10.1007/978-3-031-36644-4_8},
	language = {en},
	urldate = {2025-03-10},
	booktitle = {Machine {Learning} in {Modeling} and {Simulation}: {Methods} and {Applications}},
	publisher = {Springer International Publishing},
	author = {Dar, Zulkeefal and Baiges, Joan and Codina, Ramon},
	editor = {Rabczuk, Timon and Bathe, Klaus-Jürgen},
	year = {2023},
	doi = {10.1007/978-3-031-36644-4_8},
	pages = {297--339},
}

@article{chatterjee_introduction_2000,
	title = {An introduction to the proper orthogonal decomposition},
	volume = {78},
	issn = {0011-3891},
	url = {https://www.jstor.org/stable/24103957},
	number = {7},
	urldate = {2025-03-17},
	journal = {Current Science},
	author = {Chatterjee, Anindya},
	year = {2000},
	note = {Publisher: Temporary Publisher},
	pages = {808--817},
}

@misc{wulfinghoff_e3c_2025,
	title = {{E3C} for {Computational} {Homogenization} in {Nonlinear} {Mechanics}},
	url = {http://arxiv.org/abs/2501.13631},
	doi = {10.48550/arXiv.2501.13631},
	urldate = {2025-03-17},
	publisher = {arXiv},
	author = {Wulfinghoff, Stephan and Hauck, Jan},
	month = jan,
	year = {2025},
	note = {arXiv:2501.13631 [physics]},
}

@article{wulfinghoff_statistically_2024,
	title = {Statistically compatible hyper-reduction for computational homogenization},
	volume = {420},
	issn = {00457825},
	url = {https://linkinghub.elsevier.com/retrieve/pii/S0045782523008678},
	doi = {10.1016/j.cma.2023.116744},
	language = {en},
	urldate = {2025-03-17},
	journal = {Computer Methods in Applied Mechanics and Engineering},
	author = {Wulfinghoff, Stephan},
	month = feb,
	year = {2024},
	pages = {116744},
}

@article{wulfinghoff_model_2018,
	title = {Model order reduction of nonlinear homogenization problems using a {Hashin}–{Shtrikman} type finite element method},
	volume = {330},
	issn = {0045-7825},
	url = {https://www.sciencedirect.com/science/article/pii/S0045782517306904},
	doi = {10.1016/j.cma.2017.10.019},
	urldate = {2025-03-17},
	journal = {Computer Methods in Applied Mechanics and Engineering},
	author = {Wulfinghoff, Stephan and Cavaliere, Fabiola and Reese, Stefanie},
	month = mar,
	year = {2018},
	pages = {149--179},
}

@article{cavaliere_efficient_2020,
	title = {Efficient two–scale simulations of engineering structures using the {Hashin}–{Shtrikman} type finite element method},
	volume = {65},
	issn = {1432-0924},
	url = {https://doi.org/10.1007/s00466-019-01758-4},
	doi = {10.1007/s00466-019-01758-4},
	language = {en},
	number = {1},
	urldate = {2025-03-18},
	journal = {Computational Mechanics},
	author = {Cavaliere, Fabiola and Reese, Stefanie and Wulfinghoff, Stephan},
	month = jan,
	year = {2020},
	pages = {159--175},
}

@article{langevin_sur_1905,
	title = {Sur la théorie du magnétisme},
	volume = {4},
	issn = {0368-3893},
	url = {http://www.edpsciences.org/10.1051/jphystap:019050040067800},
	doi = {10.1051/jphystap:019050040067800},
	language = {fr},
	number = {1},
	urldate = {2025-03-18},
	journal = {Journal de Physique Théorique et Appliquée},
	author = {Langevin, P.},
	year = {1905},
	pages = {678--693},
}

@misc{wulfinghoff_homogenization_2025,
	address = {Rochester, NY},
	type = {{SSRN} {Scholarly} {Paper}},
	title = {Homogenization in {Hyperelasticity} {Using} {Empirically} {Corrected} {Cluster} ({E3C}) {Hyper}-{Reduction}},
	url = {https://papers.ssrn.com/abstract=5187007},
	doi = {10.2139/ssrn.5187007},
	language = {en},
	urldate = {2025-03-26},
	publisher = {Social Science Research Network},
	author = {Wulfinghoff, Stephan},
	month = mar,
	year = {2025},
}

@article{schenk_pardiso_2001,
	series = {I. {High} {Performance} {Numerical} {Methods} and {Applications}. {II}. {Performance} {Data} {Mining}: {Automated} {Diagnosis}, {Adaption}, and {Optimization}},
	title = {{PARDISO}: a high-performance serial and parallel sparse linear solver in semiconductor device simulation},
	volume = {18},
	issn = {0167-739X},
	shorttitle = {{PARDISO}},
	url = {https://www.sciencedirect.com/science/article/pii/S0167739X00000765},
	doi = {10.1016/S0167-739X(00)00076-5},
	number = {1},
	urldate = {2025-04-23},
	journal = {Future Generation Computer Systems},
	author = {Schenk, Olaf and Gärtner, Klaus and Fichtner, Wolfgang and Stricker, Andreas},
	month = sep,
	year = {2001},
	pages = {69--78},
}
\end{document}